\documentclass[usenatbib]{mn2e}
\usepackage{graphicx}
\usepackage{natbib}
\usepackage{amssymb}
\usepackage{aas_macros}

\title[Compact groups of galaxies]{Compact groups in theory and practice -
I. The spatial properties of compact groups}

\author[McConnachie et al.] {Alan W. McConnachie$^1$, Sara L. Ellison$^1$, David R. Patton$^{2,3}$\\
$^1$Department of Physics and Astronomy, University of Victoria, Victoria, B.C., V8P 1A1, Canada\\
$^2$Department of Physics and Astronomy, Trent University, 1600 West Bank Drive, Peterborough, ON, K9J 7B8, Canada\\
$^3$Visiting Researcher, Department of Physics and Astronomy, University of Victoria, Victoria, B.C., V8P 1A1, Canada
}

\begin{document}

\maketitle

\begin{abstract}
We use a mock galaxy catalogue based upon the Millennium Run
simulation to investigate the intrinsic spatial properties of compact
groups of galaxies. We find that approximately 30\,\% of galaxy
associations identified in our mock catalogue are physically dense
systems of four or more galaxies with no interlopers, approximately
half are close associations of 2, 3 or 4 galaxies with one or more
interlopers, and the remainder are not physically dense (projections
of looser groups and physically unassociated galaxies). Thus the
effect of interloping galaxies is significant. However, we find that
genuine compact groups are preferentially brighter and more isolated
than those with interlopers; by increasing the required minimum
surface brightness of a group from the canonical value of $\mu_e =
26$\,mags\,arcsec$^{-2}$ to $\mu_e = 22$\,mags\,arcsec$^{-2}$, we can
increase the proportion of genuinely compact systems identified with
no interlopers from $29$\,\% to 75\%. Of the genuine compact groups
identified, more than half consist of a single dark matter halo with
all the member galaxies deeply embedded within it. In some cases,
there are other galaxies which share the same halo (typically with 
mass $\sim 10^{13}\,h^{-1}$\,M$_\odot$) but which are not
identified as being members of the compact group. This implies that
compact groups are associated with group environments, some or all
members of which are in the compact group. For those compact groups
where all galaxies are in the same halo, the three-dimensional
velocity dispersion of the compact group correlates broadly with the
virial velocity of the dark matter halo. However, the scale-size of
the group - and hence the fraction of the halo mass which the group
samples - is completely uncorrelated with the properties of the dark
matter halo. This means that masses derived under the simple
assumption of virial equilibrium using the observed velocity
dispersions and sizes of compact groups give incorrect estimates of
the true mass of the underlying dark matter.
\end{abstract}

\begin{keywords}
methods: statistical - catalogues - surveys - galaxies: general - galaxies: interactions
\end{keywords}

\section{Introduction}

In broad terms, a compact group (CG) of galaxies is a group of
galaxies in which the typical inter-galactic separation is of order
the scale of the galaxies. CGs were first quantitatively defined by
\cite{rose1977} and \cite{hickson1982}, although their original
identification significantly predates these studies. Some of the most
famous examples of CGs include Stephan's Quintet
(\citealt{stephan1877}) and Seyfert's Sextet
(\citealt{seyfert1948}). The advent of the first large area surveys,
especially the Palomar Observatory Sky Survey (POSS), greatly
increased the pace of discovery of CGs. CGs make up some of the
objects identified by \cite{vorontsovvelyaminov1959},
\cite{vorontsovvelyaminov1977} and \cite{arp1966}, in their catalogues
of interacting galaxies. Studies initiated by \cite{shakhbazian1973}
resulted in the identification of a few hundred `compact groups of
compact galaxies'. Later studies by \cite{hickson1977} and
\cite{heiligman1980} identified CGs using qualitative selection
criteria based upon their morphological appearance.

The most famous and well studied catalogue of CGs is the Hickson
Compact Groups (HCGs; \citealt{hickson1982}) which were identified
from the POSS using quantitative criteria based on the number and flux
density of galaxies in a group (see Section~2.1 for a discussion of
the selection criteria). The original catalogue consists of 451
galaxies in 100 different groups. Since its creation, several other
catalogues of CGs have been compiled, including (but not limited to)
the Digitized POSS CG catalog (\citealt{iovino2003}), the Southern CG
catalogue (\citealt{prandoni1994,iovino2002}), CGs in the UZC galaxy
catalogue (\citealt{focardi2002}), the CfA2 redshift survey
(\citealt{barton1996}), the Las Campanas redshift survey
(\citealt{allam2000}), and the SDSS Commissioning Data
(\citealt{lee2004}). 

When redshifts were measured for the members of Stephan's Quintet and
Seyfert's Sextet, it was found that not all of the galaxies were at
the same redshift.  Indeed, the physical nature of CGs has been a
topic of debate for many decades (see, e.g.,
\citealt{burbidge1961}). \cite{mamon1986} argued that, instead of
being physically dense systems, roughly half of all CGs were
line-of-sight alignments of galaxies within looser groups. However,
\cite{hickson1988b} refuted this and found the probability of such an
occurrence to be $\sim 100$ smaller than predicted by
\cite{mamon1986}. More recently, \cite{hernquist1995} have suggested
that CGs may be the result of viewing a cosmological filament
end-on. \cite{ponman1996} have surveyed the HCGs with ROSAT and
implied that hot intra-group gas is present in $\sim 75\,\%$ of the
systems. The presence of such gas is strong evidence that the majority
of CGs are physically dense, even if some of their members are
interlopers.

The apparent high densities of CGs combined with their generally low
velocity dispersions (eg. \citealt{hickson1992}), make them an
intriguing environment in which to study galaxy evolution. Mergers and
interactions should be commonplace, and CGs are potentially an ideal
laboratory to study these processes. Calculations by
\cite{carneval1981} and \cite{barnes1989} suggest that CG galaxies
should merge together on a timescale equal to the crossing time of the
group, typically estimated to be of order $1$\,Gyr. If this is true,
then it suggests that CGs must be continually forming in order to
compensate for the groups lost through the merging process
(\citealt{diaferio1994}). However, \cite{governato1991} show that, for
the right choice of initial conditions, CGs can potentially last for
as long as $9$\,Gyrs before merging into a single
remnant. \cite{athanassoula1997} show that CGs can survive for much
longer than a Hubble time if they are embedded in massive dark matter
haloes.

The selection criteria for CGs inevitably favor the extrema of the
population, or those whose orientations to our line of sight are
favourable. Thus, despite their existence being known for many years,
and despite large amounts of observational effort in cataloging the
properties of the HCGs in particular, the physical nature of CGs and
their role in the galactic evolution process is still not clear.  For
example, how dense is the CG environment? How many contain
interlopers? Are they dynamically stable? Are they a transient stage
in the evolution of looser groups? Do they represent a distinct class
of galaxy environment, or are they the extreme end of a larger class?
Are they the site of the formation of early-type galaxies? What is
their relation to so-called fossil groups? What effect does the CG
environment have on the members of the group?

This is the first in a series of papers where we statistically examine
the properties of CGs and their member galaxies based on large samples
of CGs identified in cosmological simulations and observational
surveys. We begin this series by examining the selection criteria of
HCGs to determine the three dimensional reality of these systems and
the effect of interlopers. This has been a long disputed argument in
the field. The advent of large cosmological simulations and the
ability to make realistic mock galaxy catalogues means fresh insight
can now be obtained. For the first time, we can determine on a
statistical basis how many of the HCGs are likely to be genuinely
compact systems and explore their spatial properties.

Section~2 describes the Hickson selection criteria and the use of the
mock galaxy catalogue. In Section~3, we apply Hickson's criteria to
this catalogue and examine the compactness of the galaxy associations
identified. In Section~4, we optimise the selection criteria for CGs
in photometric surveys to minimise the effect of interlopers. In
Section~5, we take advantage of the cosmological simulations to
examine the halo properties of the CGs and their relation to
observable quantities. We defer the majority of the discussion of our
results until Section~6.

\section{Preliminaries}

\subsection{The Hickson criteria}

\cite{hickson1982} define a CG as a set of galaxies in a photometric
survey with projected properties such that

\begin{enumerate}
\item{$N\left(\Delta\,m = 3\right) \geq 4$;}
\item{$\theta_N \geq 3\,\theta_G$;}
\item{$\mu_e \leq 26.0$.}
\end{enumerate}

\noindent $N\left(\Delta\,m = 3\right) $ is the number of galaxies
within 3\,magnitudes of the brightest galaxy and $\mu_e$ is the effective
surface brightness of these galaxies in magnitudes per square arcsec
where the total flux of the galaxies is averaged over the smallest
circle which contains their geometric centres, and has an angular
diameter $\theta_G$. $\theta_N$ is the angular diameter of the largest
concentric circle which contains no additional galaxies in the magnitude
range of the group or brighter. All magnitudes and surface
brightnesses are measured in the $r-$band. We hereafter refer to these
three criteria collectively as the `Hickson criteria'.

The identification of groups of only $\sim 4$ galaxies from large
photometric surveys where the only data available are $x, y$ projected
positions and apparent magnitudes is a difficult problem. The Hickson
criteria were originally constructed in an attempt to quantify what
had up until then been identified by eye, and are inevitably
subjective. Criterion (i) ensures that the CG has enough members
to constitute a group. A magnitude range is imposed to ensure all
members are of broadly equivalent masses. Without it, a galaxy and its
satellites (such as the Milky Way and its entourage of dwarf
spheroidals) could be identified as a CG. Criterion (ii) attempts to
distinguish isolated groups from parts of larger structures, such as
cluster cores.  Criterion (iii) is, to first order, a distance
independent means of defining compactness.

These criteria have been subject to some controversy since Hickson
first introduced them. The basic issue is that it is not clear what
the physical, three-dimensional reality of the identified galaxy
associations are, given that the criteria are applied to projected,
two-dimensional datasets. In particular, it is difficult to assess how
homogeneous the identified galaxy associations are in terms of their
physical compactness and environment, and the contribution from
interlopers is unknown. These are fundamental problems for any
observational study of CGs.

In what follows, we use the following terminology: `Hickson
association (HA)' describes all of the galaxy associations identified
in the mock catalogue by application of the Hickson criteria. This is
in recognition of the fact that some of these associations will be
genuinely compact configurations in three dimensions, some will be
projections of looser groups and some may be entirely spurious
line-of-sight chance alignments. The term `compact association (CA)'
is used to describe the subset of HAs which are truly compact in three
dimensions (we define `compact' in Section~3.1).

\subsection{The mock galaxy catalogue}

The primary goal of this study is to assess the spatial and dynamical
properties of compact galaxy groups by using simulations to relate
three dimensional group properties to those in redshift space.  The
Millennium Run simulation of \cite{springel2005} is a natural starting
point: this simulation follows the evolution of dark matter within a
cube which is 500\,$h^{-1}$\,Mpc on a side, from a redshift of 127 to
the present day.  Here, the Hubble constant is parameterised as $H_o =
100\,h\,$km\,s$^{-1}$Mpc$^{-1}$. Several galaxy catalogues have been
created by applying semi-analytical models to the output of these
simulations (\citealt{croton2006,delucia2007, bower2006}).  In
principle, any or all of these catalogues are suitable for the
three-dimensional requirements of our compact groups study.

In order to facilitate a direct comparison between simulations and
observations, we require a mock redshift survey from these galaxy
catalogues.  For this task, we use the publicly available Blaizot
Allsky catalogues available on the Millennium
website\footnote{http://www.mpa-garching.mpg.de/millennium/}.  These
mock catalogues were created by applying the Mock Map Facility (MoMaF)
code of \cite{blaizot2005} to the \cite{delucia2007} semi-analytic
galaxy catalogues. The basic idea which creates the mock catalogue is
the replication of the data cube from the simulations (which has
periodic boundary conditions) in all directions.  Evolution is built in
by using time snapshots which correspond to the look-back time of each
cube.

Five of the six available Allsky catalogs employ the random tiling
technique, which avoids spatial correlations on large scales due to
replication of the simulation cube.  We choose not to use these
catalogues, since the discontinuities which arise between adjacent
data cubes may lead to anomalous galaxy group detections.  Instead, we
use the ``Blaizot\_Allsky\_PT\_1'' catalogue, which preserves the
periodicity of the density field.  We note, however, that the
periodicity is not a concern on the small scales on which compact
galaxy groups are found.  Moreover, even if a given galaxy group is
detected more than once, it will be viewed from different vantage
points, and is therefore likely to have different observable
properties.

The mock catalogue we use consists of $\sim 5.7$\,million galaxies
brighter than the magnitude limit of $r_{AB} = 18$, where $r$ is the
SDSS red filter.  These catalogues contain the key observable
properties one would have in a flux-limited survey: namely right
ascension, declination and apparent magnitude.  In addition, much of
the associated information from the \cite{delucia2007} galaxy
catalogues is also provided, enabling a direct link to three
dimensional positions and velocities, as well as intrinsic properties
such as absolute magnitudes and stellar mass. These properties will be
looked at in detail for the compact groups we detect in the next paper
in this series (\citealt{brasseur2008}). Here, we are now in a
position to detect compact groups in a flux limited mock catalogue,
and to relate the observed spatial properties of these groups to their
counterparts in three dimensions.

The results we derive are only strictly correct for the mock galaxy
catalogue from which they are derived, and any significant change in
the input physics of the simulations on which they are based could
affect our results. However, the \cite{delucia2007} semi-analytic
model galaxy catalogues used in this paper are closely related to
those of \cite{croton2006}; the latter demonstrated that their galaxy
catalogues are broadly consistent with observations of low redshift
galaxies, including the luminosity function, the global star formation
history, the Tully-Fisher relation, and the colour-magnitude
distribution.  In addition, the clustering properties of the
\cite{croton2006} catalog appear to be well matched to observations
(\citealt{springel2005,li2007}).  

Compact groups have, by definition, small scale radii, and it is
therefore important to check that the resolution of the simulation is
sufficient to identify these systems. The spatial resolution of the
Millennium simulation is 5\,$h^{-1}$\,kpc
(\citealt{springel2005,croton2006}), set by the softening length of
the gravitational force. This resolution scale is significantly less
than the mean separation of galaxies in compact groups (Section 3.1)
and will not affect the majority of systems we identify. However,
interactions between galaxies at {\it very} small separations will not
be followed accurately.

Similarly, the mass resolution for the identification of galaxy haloes
in the simulation is important, particularly during the merging of
galaxies. When two haloes start to merge, \cite{delucia2007} follow
the merger until the substructure drops below the mass resolution
limit for the identification of (sub-)haloes. At this point, the two
haloes are assumed to fully merge after a time $t_{merge}$, which is
assumed to be twice the dynamical time. However, this timescale is
{\it ad-hoc}. Changing $t_{merge}$ could affect the number of galaxies
we identify as compact groups. To investigate whether this severely
affects our results, we calculate the distribution of sub-halo masses
for all the galaxies we identify as members of compact groups (i.e., HAs;
the halo masses are provided in the \cite{delucia2007}
catalogues). The mass resolution for the identification of
(sub-)haloes in the simulation is $\sim 1.7 \times
10^{10}\,h^{-1}\,$M$_\odot$, and we find that less than $1\,\%$ of
galaxies in HAs have haloes this small. Only 5\% of the
galaxies have haloes with mass $< 2 \times
10^{10}\,h^{-1}$\,M$_\odot$, and more than half of the haloes are
larger than the mass resolution limit by more than a factor of 10. The
halo masses of CAs (genuinely dense groups of galaxies) are higher
still (Section~5). Thus possible uncertainties associated with
$t_{merge}$ are likely to have only a small, if not negligible, effect
on this study.

Future generations of cosmological simulations will have improved
spatial and mass resolution and will be able to more accurately trace
the formation and evolution of dense groups of galaxies like compact
groups. While this study - like any based on mock catalogues - cannot
be definitive due to the many physical and computational uncertainties
inherent in simulating cosmological volumes, it is nevertheless
impressive that the simulations are of sufficient quality to start to
provide a new handle on, and new insight into, the properties of
compact groups. Our results are as robust as our current understanding
allows, and should provide a useful comparison to samples of compact
groups extracted from galaxy surveys such as the SDSS.

\section{The predicted spatial characteristics of compact groups}

\subsection{Degree of compactness}

\begin{figure}
  \begin{center}
    \includegraphics[angle=270, width=8.cm]{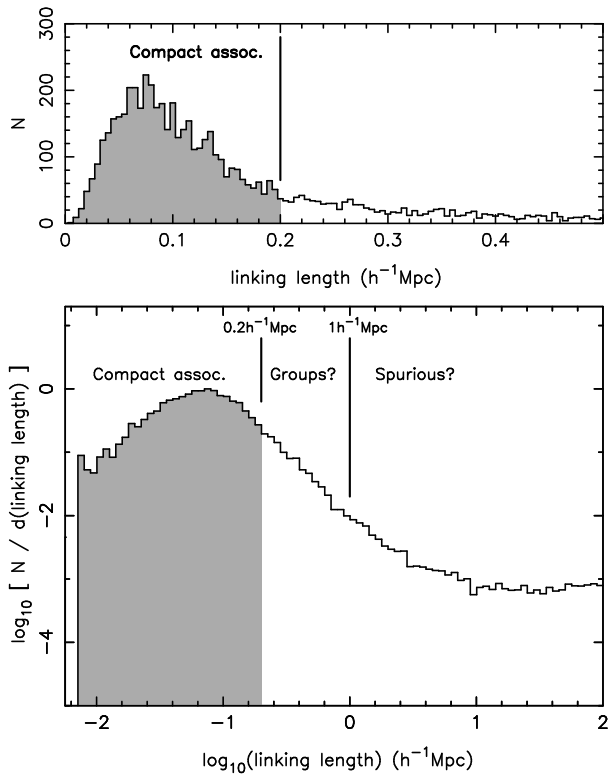}
    \caption{Distribution of linking lengths of all identified Hickson
    associations (HAs) in the mock galaxy catalogue, shown on linear
    (top panel) and log (bottom panel) scales. The Hickson criteria
    preferentially identify galaxy associations with small linking
    lengths, although there is a very significant tail of associations
    with large linking lengths. Shaded areas show all HAs with linking
    lengths less than 200\,$h^{-1}$\,kpc.}
    \label{f1}
  \end{center}
\end{figure}

\begin{figure}
  \begin{center}
    \includegraphics[angle=270, width=8.cm]{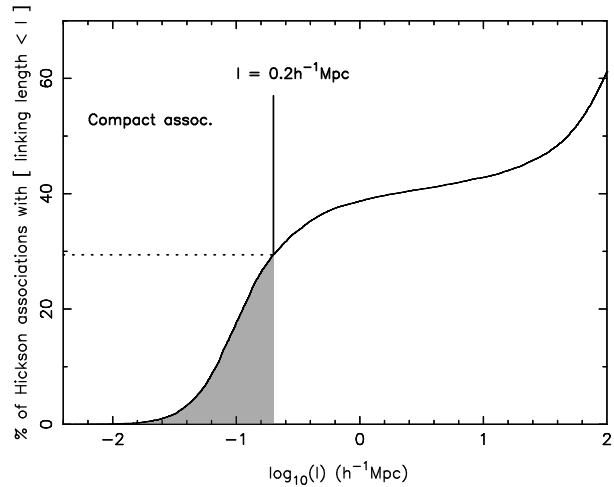}
    \caption{Percentage of Hickson associations with linking lengths
    $\le l$. Approximately $29\,\%$ of all HAs (4446 out of 15122)
    have linking lengths less than $l = 200$\,$h^{-1}$\,kpc (shaded
    area).}
    \label{f2}
  \end{center}
\end{figure}

We implement a search algorithm on the mock galaxy catalogue using the
Hickson criteria to identify all HAs. In keeping with the original
criteria, we use $r-$band apparent magnitudes to calculate surface
brightnesses. We find $15\,122$ HAs in the mock galaxy catalogue which
satisfy the Hickson criteria, consisting of a total of $64\,525$
galaxies. The benefit of this approach is that we can now cross
correlate the two dimensional, projected, properties of each galaxy in
every HA with their de-projected properties to determine the three
dimensional reality of the HAs.

The intrinsic compactness of a HA is fundamental to studies of compact
groups. The effective surface brightness, $\mu_e$, is used in
projection as a means of characterising compactness. Since three
dimensional spatial information is available to us, we choose to use
the `linking length', $\ell$, as a quantitative and robust measure of
the three dimensional compactness of each of the 15\,122 HA identified
in our mock catalogue. This parameter is used extensively in
friends-of-friends algorithms which are commonly used to identify
galaxy groups when redshift information is available
(\citealt{huchra1982}).  For our purposes, we calculate the value of
$\ell$ for each HA in physical coordinates which corresponds to the
minimum linking length that would need to be used to ensure all
members of that HA were identified in the simulation upon application
of a friends-of-friends algorithm. The use of linking length as a
measure of compactness has the added advantage that it is independent
of the overall shape of the group: for example, it does not
discriminate between a roughly spherical group or an elongated
filamentary structure.

The distribution of three dimensional linking lengths for the
identified HAs is shown in Figure~\ref{f1}. The top panel shows the
distribution at small linking lengths ($\ell < 500$\,$h^{-1}$\,kpc) on a linear
scale whereas the lower panel shows the linking lengths for all HAs on
a log scale (where each bin has been normalised by its width).  The
scale on which a HA becomes a CA is, to some extent,
subjective. However, the top panel of Figure~\ref{f1} reveals that
Hickson's criteria preferentially identify HAs with small linking
lengths, of order a few hundred kpc or less. Despite this, there is a
tail to very large values of $\ell$, as the lower panel of
Figure~\ref{f1} makes clear. To quantify this further, Figure~\ref{f2}
shows the percentage of HAs with linking lengths less than or equal to
$\ell$.

By our definition, compact associations (CAs) have small values of
$\ell$, of order the scale of the galaxies. Looser groups have
intermediate values of $\ell$; for example, the linking length of the
brightest members of the Local Group, a relatively sparse grouping of
galaxies, is approximately $800$\,kpc. Spurious detections, caused by
line-of-sight alignments of unassociated galaxies at different
redshifts, yield very large values of $\ell$. Therefore, a linking
length $\gg 1\,h^{-1}$\,Mpc would indicate that at least some of the
galaxies in the HA are unlikely to be physically related to one
another. We have indicated approximate `boundaries' between CAs, loose
groups and spurious associations in the lower panel of
Figure~\ref{f1}.

Given the shape of the distribution of linking lengths in the top
panel of Figure~\ref{f1}, and the curve of the distribution in
Figure~\ref{f2}, $\ell = 200$\,$h^{-1}$\,kpc seems a natural division
below which groups are classed as compact. It is also approximately
the virial radius of $L\star$ galaxy haloes such as that of the Milky
Way and M31 (\citealt{klypin2002}). Since compact groups are generally
understood to be galaxy groups where the typical inter-galactic
separation is of order the scale of the galaxies themselves, we adopt
$\ell \leq 200$\,$h^{-1}$\,kpc as the appropriate scale for CAs. This
corresponds to the shaded areas of Figures
\ref{f1}~and~\ref{f2}. Defined in this way, $\sim 29$\,\% of all HAs
(4446 galaxy associations) are compact associations of 4 or more
galaxies in three dimensions.

The boundary between compact associations and looser arrangements of
galaxies at $\ell = 200$\,$h^{-1}$\,kpc is somewhat subjective and
could reasonably be set at a slightly higher value to increase the
fraction of associations we deem compact. We note, however, that
increasing the boundary by 50\,\% to $\ell = 300$\,$h^{-1}$\,kpc only
increases the number of HAs identified as CAs by 4\% to 33\%. This is
probably as large as the cut could reasonably be set given the shape
of the distribution of linking lengths in Figures 1 and 2, and it does
not significantly change the proportion of HAs we consider to be
compact.  The cut could be set much smaller than 200\,$h^{-1}$\,kpc,
of course, and this would reduce the number of systems which we
consider as CAs quite dramatically.

\subsection{Interloping groups and interlopers within groups}

\begin{figure}
  \begin{center}
    \includegraphics[angle=270, width=8.cm]{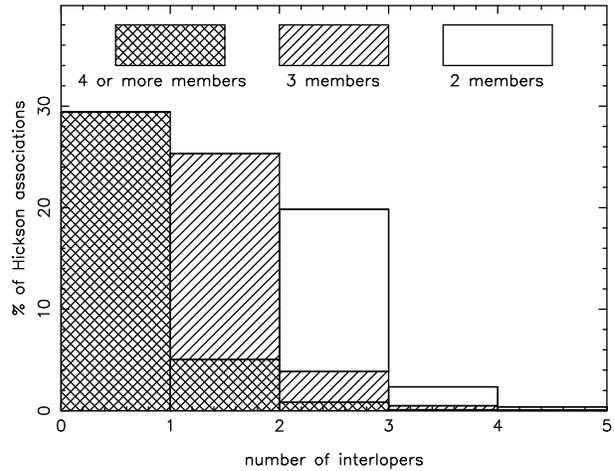}
    \caption{Percentage of Hickson associations (HAs) which satisfy
    our compactness criteria for compact associations (CAs) (linking
    length $\le 200$\,$h^{-1}$\,kpc), allowing for the presence of interloping
    galaxies. Approximately $35\,\%$ of HA consist of a compact
    configuration of 4 or more galaxies, an additional $24$\,\%
    consist of 3 galaxies in a compact configuration, and an
    additional $18$\,\% of HAs consist of a pair of galaxies separated
    by less than 200\,$h^{-1}$\,kpc. However, many of these `groups' would
    otherwise fail to satisfy the selection criteria for HAs were it
    not for the interloper members.}
    \label{f4}
  \end{center}
\end{figure}

Figure~\ref{f2} shows that $\sim 71$\,\% of HAs do not have small
enough values of $\ell$ to be considered as CAs based on our
definition of compactness. This fraction is quite large, and would
suggest that the Hickson criteria do not efficiently identify compact
arrangements of galaxies. However, while some of the HAs with large
$\ell$ will be completely spurious line-of-sight alignments of
galaxies (`interloping groups'), another possibility is that {\it
some} of the galaxies within the HA form a compact arrangement, and
the large $\ell$ value is the result of a minority of physically
unassociated galaxies which happen to lie along the same line of sight
at a very different distance (`interloping galaxy/galaxies'). There are
famous examples of this; for example, four out of the five galaxies
which form Stephan's Quintet are at the same redshift, whereas one is
not.

To gauge how this affects our results, we calculate how many of the
HAs with $\ell > 200$\,$h^{-1}$\,kpc can be decomposed into a CA ($\ell \le
200$\,$h^{-1}$\,kpc) of 4 or more members and one or more interlopers. The
results are shown as a percentage of the total number of HAs in
Figure~\ref{f4}. The first bin is for all CAs with 4 or more galaxies
and no interlopers, and corresponds to the $\sim 29\%$ discussed in
Section~3.1. The remaining bins show the fraction of HAs which can be
decomposed into a CA with 4 or more genuine members plus one or more
interlopers (cross-hatched histograms). In total, the percentage of HAs
identified as CAs increases by $\sim 6\,\%$. Of these $6\,\%$, virtually none
have more than 2 interloping galaxies. Thus just over one-third
($35\,\%$) of identified HAs are compact arrangements of 4 or more
galaxies in three dimensions, allowing for the presence of
interlopers.

HAs are defined to have at least four members. However, given that
interlopers are inevitably present in any observation sample of galaxy
groups or clusters, we relax the requirement that 4 member galaxies
are required to define a CA. Instead, we require a minimum of only 3
members, still with $\ell \le 200$\,$h^{-1}$\,kpc (hatched histograms
in Figure ~\ref{f4}). By doing this, we find an additional $\sim 24\%$
of HAs can be decomposed into a CA plus interlopers. This means
that $\sim 59\,\%$ of identified HAs consist (at least in part) of a
compact arrangement of $\geq 3$ galaxies.

Finally, we determine that an additional 18\,\% of HAs consist of a
pair of galaxies separated by less than 200\,$h^{-1}$\,kpc with at
least 2 interlopers (blank histograms in Figure~\ref{f4}).  The number
of completely `spurious' associations identified by the Hickson
criteria (that is, those where none of the galaxies are closer than
200\,$h^{-1}$\,kpc to any other) is therefore less than one in four;
77\,\% of HAs consist of 2 or more galaxies which are physically
associated on scales we consider `compact'.

It is important to emphasise that most of the compact arrangements of
galaxies with one or more interlopers would otherwise fail to meet the
criteria which define a HA were it not for the interlopers; they would
either not have enough members or have too low a projected surface
brightness to qualify.

Figure~4 summarises the results of this section. The finding that most
($77\,\%$) HAs are physically close associations of galaxies is
clearly significant, as is the result that most of these physically
close groupings usually consist of one or more interlopers. Of the
64\,525 galaxies which make up all the identified HAs, 25\,497
galaxies ($\sim 40$\,\%) are not physically close ($< 200\,h^{-1}$\,kpc) to
other galaxies. The effect of contamination from interlopers is
clearly significant, and we will discuss this in more detail in
Section~6.

\begin{figure}
  \begin{center}
    \includegraphics[angle=270, width=8.cm]{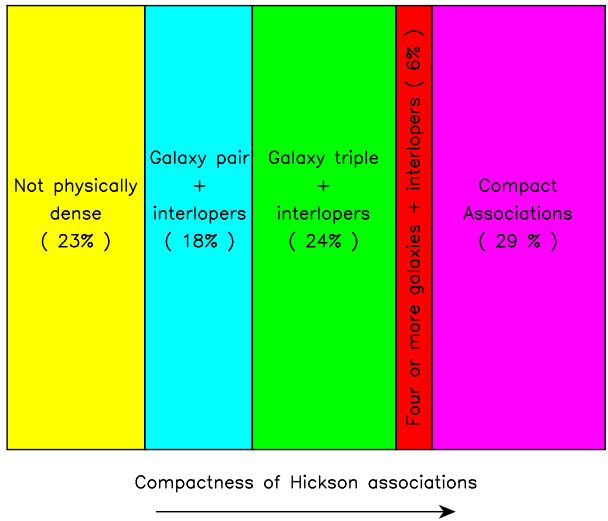}
    \caption{Summary of the spatial properties of the galaxy
    associations identified using Hickson's criteria for compact
    groups. The area of the rectangles are proportional to the number
    of such systems in our sample of 15\,122 Hickson associations.}
  \label{f8}
  \end{center}
\end{figure}

\section{Optimal selection criteria}

\begin{figure*}
  \begin{center}
    \includegraphics[angle=270, width=14.cm]{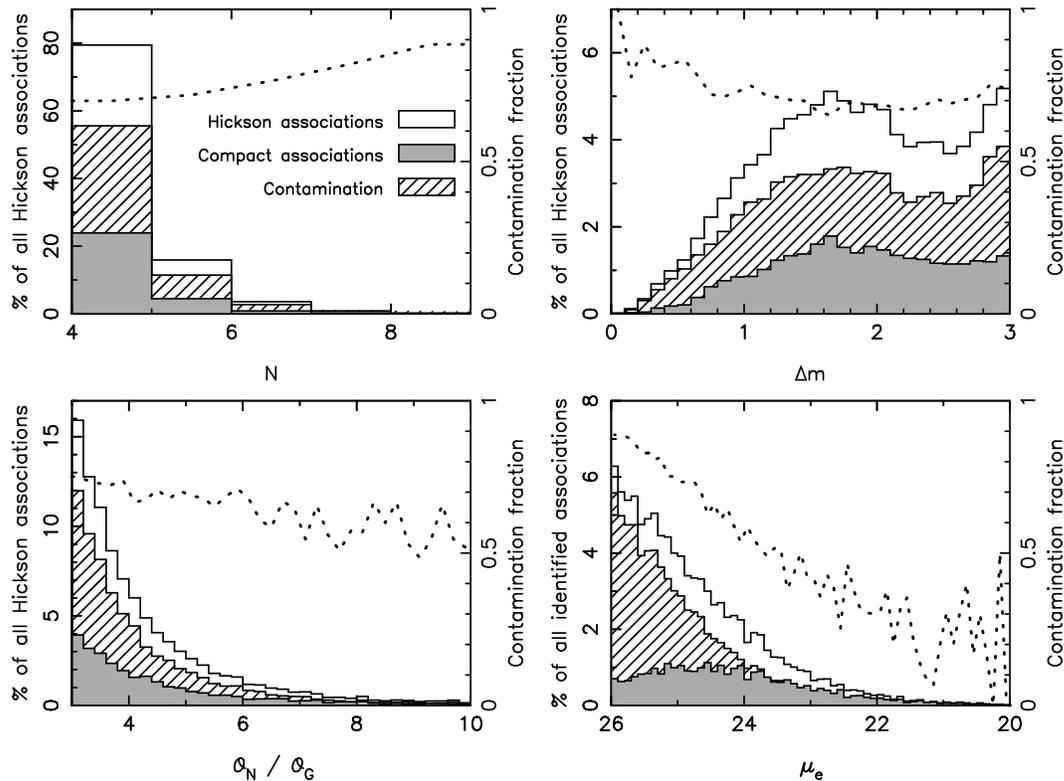}
    \caption{Distribution of observable parameters in the Hickson
    criteria (number of galaxies, $N$; apparent $r-$band magnitude
    range of members, $\Delta m$; projected distance to nearest
    non-member galaxy, $\theta_N/\theta_G$; effective surface
    brightness, $\mu_e$). Blank histograms show all Hickson
    associations (HAs), shaded histograms show compact associations
    (CAs), and hatched histograms show contamination (where we define
    all HAs which are not CAs as contamination). The dotted line in
    each panel shows the contamination level, expressed as a
    fraction of the number of HAs, as a function of each
    parameter (right vertical axis).}
  \label{f7}
  \end{center}
\end{figure*}

\begin{table}
\begin{tabular}{rcccrcc}
\hline
$N$ & \% CAs & \% HAs && $\Delta\,m$ & \% CAs & \% HAs\\
\hline
4 & 100 & 29 && 3.0 & 100 & 29 \\
5 & 19 & 27 && 2.5 & 79 & 30 \\
6 & 4 & 22 && 2.0 & 57 & 29 \\
7 & 1 & 17 && 1.5 & 31 & 27 \\
\hline
$\theta_N/\theta_G$ & \% CAs & \% HAs && $\mu_e$ & \% CAs & \% HAs\\
\hline
3 & 100 & 29 && 26 & 100 & 29 \\
4 & 51 & 34 && 25 & 72 & 43 \\
5 & 29 & 37 && 24 & 39 & 56 \\
6 & 19 & 41 && 23 & 16 & 67 \\
\hline
\end{tabular}
\caption{The effect of varying the observable parameters for the
various selection criteria of HCGs. The standard Hickson criteria have
$N \ge 4$, $\Delta\,m \le 3$\,magnitudes, $\mu_e \le
26.0$\,mags\,arcsec$^{-2}$ and $\theta_N/\theta_G \ge 3$. This
identifies 4446\,CAs in our catalogue (shown as \% of CAs $= 100$ in
the table) which constitute 29\,\% of the HAs identified. Each of the
parameters is varied independently in this table, with the other
parameters set to their standard values.}
\label{t1}
\end{table}

Given the significant number of interloping galaxies identified, we
now examine whether it is possible to alter the Hickson criteria to
reduce the `contamination' levels, so that fewer interloping groups
and interloping galaxies are identified. The associations we are
interested in are the physically compact associations (CAs) of 4 or
more galaxies with no interlopers. This is a much more homogeneous
population than the overall sample of HAs, and are most analogous to
the systems that the Hickson criteria were designed to identify. While
pairs and triplets of galaxies are interesting environments in their
own right, they are clearly not what the Hickson criteria were
attempting to identify, insofar as they do not have enough members to
satisfy Hickson's definition of a group. We therefore treat {\it all}
galaxies which do not belong to a CA as our 'contamination'.

We want to stay as true to the original definition of (H)CGs as
possible. Therefore, we look to fine-tune the selection criteria of
\cite{hickson1982} to reduce contamination rather than create a
completely new set of criteria. The selection criteria of
Hickson detailed in Section~3.1 have 4 observable parameters whose
values can be varied; these are $N$ (number of member galaxies),
$\Delta\,m$ ($r-$band magnitude range of members), $\mu_e$ (effective
surface brightness of group), and $\theta_N/\theta_G$ (degree of
isolation in projection). The distribution of each of these four
observables for all HAs, CAs and contamination identified in our mock
catalogue are shown in Figure~\ref{f7} as blank, shaded and hatched
histograms, respectively.

Figure~\ref{f7} shows the observed properties of the CAs compared to
the contaminants, and as such allows us to explore how variations in
the cuts for the observed parameters increase or decrease the
contamination rate. The dotted line in each panel shows how the
contamination level, expressed as a fraction of the number of HAs,
changes as a function of each parameter (right vertical
axis). Table~\ref{t1} shows the net effect of varying each of the four
parameters in turn, with the other three parameters set to their
canonical values given in Section~3.1. For each parameter, the first
column gives the value it is set to, the second column gives the
number of CAs retrieved (shown as a percentage of the number of CAs
obtained on using the standard values for all the parameters, 4446)
and the third column gives the number of CAs retrieved as a percentage
of the number of HAs retrieved.

Table~\ref{t1} shows that the only observable parameter whose value
does not appear to significantly change the contamination level is
$\Delta\,m$. Making $\Delta\,m$ smaller by a couple of magnitudes has
no significant change on the contamination level of the sample
retrieved. However, as the dotted line in the upper-right panel of
Figure~\ref{f7} makes clear, the contamination rate at large
$\Delta\,m$ ($2.5 - 3$\,magnitudes) and small $\Delta\,m$ ($0 -
0.5$\,magnitudes) is slightly higher than at intermediate values. Most
of the galaxy associations are identified at intermediate values,
however, and so the overall contamination levels for the cuts listed in
Table~1 do not change significantly.

More interestingly, Figure~\ref{f7} and Table~\ref{t1} reveal that the
level of contamination of more populated groups is higher than less
populated groups. This is unsurprising, given that it is more likely
that a HA with many members has an interloper as a member than does a
HA with fewer members. However, in terms of reducing contamination, it
is not worth changing the requirements on how many galaxies define a
group: although contamination is higher for more populated groups,
there are very few groups which have more than 4 or 5 members in the
first place.

The requirement that CAs should appear isolated can be used to reduce
the level of contamination. As Table~1 and Figure~6 make
clear, CAs preferentially appear more isolated on the sky than the
contaminants and selecting groups which are increasingly distant from
their nearest neighbour reduces the contamination rate. The fraction
of CAs present in a sample of HAs can increase from 29\% using the
standard criteria to 41\% if we require $\theta_N \ge 6\,\theta_G$. A
price is paid, however, since the sample size becomes less than
one-fifth of its original size.

By far the most striking feature of Figure~5 and Table~1 is the
distribution of $\mu_e$ in the lower right panel for the CAs compared
to the contaminants. The number of contaminants with a given surface
brightness increases steeply towards the faint end. In contrast,
however, the number of CAs with lower and lower surface brightness
levels off and starts to decline past a surface brightness of $\mu_e
\simeq 24 - 25$\,mags\,arcsec$^{-2}$. This is to be expected: CAs need
to have at least 4 galaxies in a relatively compact configuration in order to
be identified. Therefore, the effective surface brightness of a CA
(which is to first order distance independent) cannot be arbitrarily
faint since the galaxies must all reside within a finite volume. There
is no such physical restriction on the contaminants. The distribution
of $\mu_e$ for CAs will therefore peak in some broad range, while that
for the contamination will not.

Given the different shapes of the distribution of $\mu_e$ for CAs and
the contaminants, adopting different cuts in $\mu_e$ has a dramatic
effect on the contamination level of HAs, as the dotted line in the
lower right panel of Figure~\ref{f7} and Table~1 demonstrates. By
requiring HAs to have a minimum effective surface brightness of
25\,mags\,arcsec$^{-2}$ instead of 26\,mags\,arcsec$^{-2}$, we
increase the fraction of CAs identified to 43\% from 29\%. The number
of CAs identified is then 72\,\% of the original number. Requiring a
minimum surface brightness of 24\,mags\,arcsec$^{-2}$ means that over
half of the identified HAs are CAs. At $\mu_e =
23$\,mags\,arcsec$^{-2}$, the proportion is 67\%. While not listed in
Table~1, this trend continues to $\mu_e = 22$\,mags\,arcsec$^{-2}$,
where the proportion of CAs is 75\%. By this point, however, far fewer
CAs are identified. Clearly, searches for CAs which have higher
surface brightness cuts have significantly reduced contamination
levels.

\section{Halo occupation and dynamics}

Despite being identified using only projected information, the Hickson
criteria do a remarkable job at retrieving dense groupings of
galaxies. With only moderate refinement of the selection criteria we
have shown that we can improve on these criteria further to minimise
interlopers. We now ask how the observed dynamical properties of the
member galaxies relate to the properties of the underlying dark matter.

\subsection{Haloes versus galaxies}

\begin{figure}
  \begin{center}
    \includegraphics[angle=270, width=8.cm]{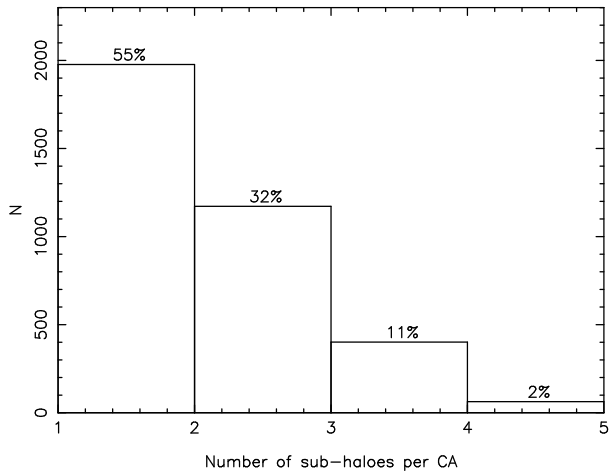}
    \caption{Histogram showing the number of distinct haloes in
    each identified compact association with 4 members. In over half
    of these compact associations all the member galaxies occupy the
    same halo, confirming their physical association and
    co-evolution.}
  \end{center}
\end{figure}

Figure~6 shows the number of separate haloes associated with each CA
previously identified in Section~3. To simplify the analysis, we use
only CAs with 4 members. Over half of the CAs consist of only one
halo, with all four galaxies resident within that same halo. Of the
rest, only a handful contain four individual haloes (one per
galaxy). There is clearly not a one-to-one correspondence between
haloes and galaxies in CAs, and most galaxies in CAs are found within
the same halo. This is a strong affirmation of the physical
association of galaxies in a CA environment, and highlights their
obvious inter-dependence and co-evolutionary status.

The \cite{delucia2007} catalogue records whether a galaxy is at the
center of its halo or if it is a satellite. In 99\,\% of the cases
where all member galaxies are within the same dark matter halo, one
galaxy is recorded as being at the center of that halo. In the
remaining 1\% of cases, all four galaxies are listed as satellites.

\subsection{Halo occupation/environment of compact groups}

\begin{figure}
  \begin{center}
    \includegraphics[angle=270, width=8.cm]{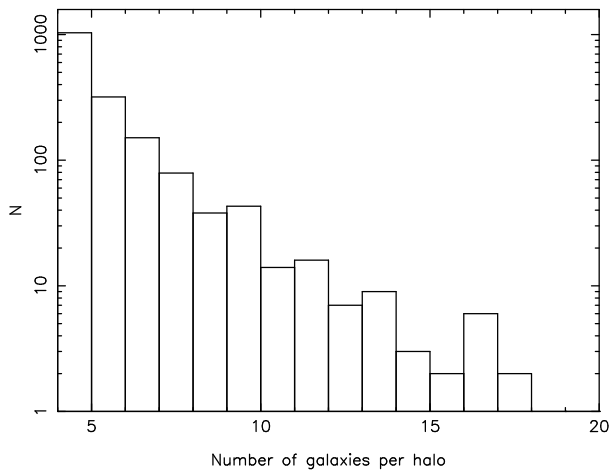}
    \caption{Number of galaxies in each halo known to contain a CA
    with four members. We consider only those galaxies within three
    magnitudes of the brightest galaxy in the CA, to keep in line with
    the original Hickson criteria. Most haloes only contain these four
    galaxies, although some haloes contain a handful of other galaxies
    which are not part of the CA. In more than half of cases, CA
    galaxies generally appear to dominate the halo in which they are
    found, although some are associated with larger groups of
    typically 5 - 10 galaxies in total.}
  \end{center}  
\end{figure}

For those CAs with four members all embedded in the same dark matter
halo, we now ask if these are the only four galaxies in the halo, or
if there are others embedded in the same halo but which do not form
part of the CA. This probes the environment of CAs; if CAs are
embedded in larger structures, such as groups or clusters, then we
expect many other galaxies to be resident in the same haloes.

Figure~7 shows the halo occupation distribution for those haloes we
know to contain a CA of four galaxies. Only those galaxies within
three magnitudes of the brightest galaxy in the CA are considered, to
keep in line with the original Hickson criteria. The large majority of
these haloes only contain the four galaxies of the CA, with the
remainder containing, at most, a handful of extra galaxies. Therefore,
most CAs dominate the haloes in which they are found, although some
are associated with larger groups of typically $5 - 10$ galaxies in
total. We have not probed the environment of the halo, but such an
analysis is beyond the scope of this paper. We note, however, that it
is unlikely that many CAs are associated with galaxy clusters, since
in this scenario we would expect many other galaxies to share the same
halo as the CA.

\subsection{Halo mass and group dynamics}

\begin{figure}
  \begin{center}
    \includegraphics[angle=270, width=8.cm]{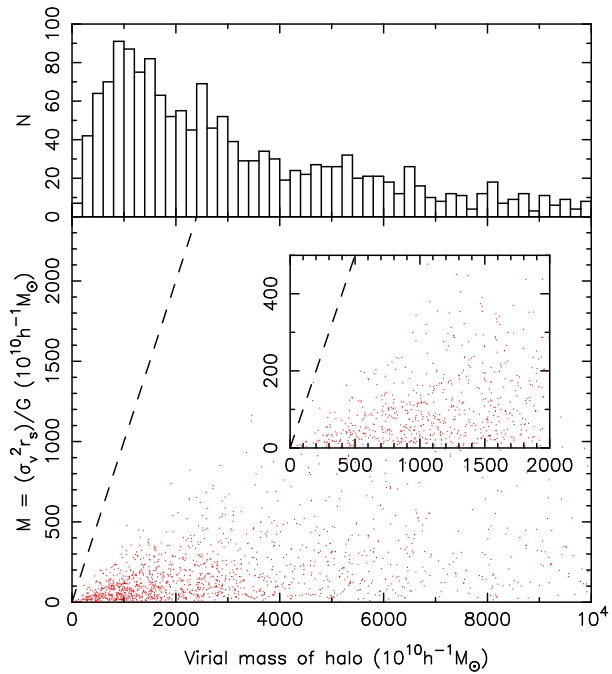}
    \caption{Top panel: histogram of the halo virial mass distribution
    for CAs with 4 members all embedded in the same dark matter
    halo. Bottom panel: the halo virial mass versus masses derived
    using the velocity dispersion and scale radius for the CAs. The
    dashed line shows a one-to-one correspondence. The inset panel
    shows the same distribution on smaller scales. The derived masses
    are a gross underestimate of the virial mass of the halo, and the
    two are uncorrelated with one another.}
  \end{center}
\end{figure}

\begin{figure}
  \begin{center}
    \includegraphics[angle=270, width=8.cm]{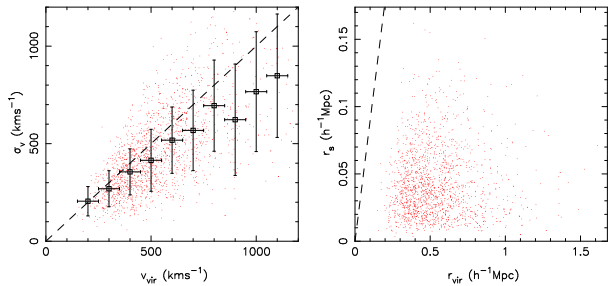}
    \caption{Left panel: halo virial velocity, $v_{vir}$, versus
    three-dimensional velocity dispersion, $\sigma_v$, for CAs with 4
    members all embedded in the same dark matter halo. The dashed line
    shows a one-to-one correspondence. The points represent the median
    values of $\sigma_v$ in 100\,kms$^{-1}$ bins in
    $v_{vir}$. Horizontal error bars show the width of the bin, and
    vertical error bars show the dispersion in $\sigma_v$ in each
    bin. The two quantities correlate strongly with one another, but
    the scatter in $\sigma_v$ is significant, of order $100 -
    300$\,km\,s$^{-1}$. Right panel: The virial radius of the halo,
    $r_{vir}$, versus scale radius of the CA ($r_s$; Equation~1).  The
    dashed line shows a one-to-one correspondence. The galaxies are
    deeply embedded in the dark matter halo, and the two quantities
    are uncorrelated with one another.}
  \end{center}
\end{figure}

The galaxies in CAs which occupy only one halo might be viable tracers
of the masses of these haloes. Indeed, the velocity dispersion of
compact groups is usually used to imply a mass for the group, but
small number statistics and the necessary assumption of virial
equilibrium means that it is not clear that this method will work for
these systems.

The top panel of Figure~8 shows the halo virial mass distribution for
all those CAs with 4 member galaxies which all reside in the same
halo. The typical halo mass is $\sim 10^{13}$\,$h^{-1}$\,M$_\odot$ - normal for
a galaxy group - although there are a considerable number more massive
than this.

The three dimensional velocity dispersion, $\sigma_v$, of each of
these CAs is plotted as a function of the virial velocity of the host
halo, $v_{vir}$, in the left panel of Figure~9. Typical velocity
dispersions are of order $300 - 500$kms$^{-1}$. The dashed line shows
a one-to-one correspondence between $v_{vir}$ and $\sigma_v$. The
squares show the median velocity dispersion in 100kms$^{-1}$ bins in
virial velocity. The horizontal error bars show the width of the bin,
and the vertical error bars show the standard deviation of the
velocity dispersions in that bin. Clearly, the velocity dispersion of
the group broadly correlates with the virial velocity of the host
halo. The velocity dispersion is systematically lower than the virial
velocity, and the spread about the median is very significant, of order
$100 - 300$\,km\,s$^{-1}$. This means that adopting the velocity
dispersion as a proxy for the dark matter halo virial velocity for any
individual system is going to introduce a significant source of
uncertainty.

To calculate the mass of the halo via the virial theorem requires a
representative scale radius for the galaxy group. Following
\cite{binney1987}, we use

\begin{equation}
r_s = \frac{1}{\Sigma_{i=1}^n \Sigma_{j<i} <\left|R_i - R_j\right|^{-1}>}~,
\end{equation}

\noindent which is a measure of the typical intergalactic separation
in the group. The second panel of Figure~9 shows $r_s$ as a function
of the virial radius of the halo, $r_{vir}$. The dashed line
represents a one-to-one correspondence between these quantities. The
scale radius of the CA is much smaller than the virial radius of the
dark matter halo, showing that the CA is deeply embedded in the
central regions of the host halo (in Section~5.1 we showed that one
galaxy in the CA is usually at the center of the halo). Further, the
degree to which the CA is embedded is independent of $r_{vir}$, as
demonstrated by the lack of a correlation between $r_{vir}$ and $r_s$.

Using the observed $\sigma_v$ and $r_s$ for the CAs, we can calculate
the mass that would be derived for the host halo upon simple
application of the virial theorem. The lower panel of Figure~8 shows
this mass in comparison to the true (virial) mass of the dark matter
halo. The dashed line shows a one-to-one correspondence. The inset
panel shows the same distribution on smaller scales. Clearly, the
estimated mass is a gross underestimate of the virial mass of the
halo, but this is to be expected; the galaxies are very embedded in
the dark matter halo and sample a smaller mass corresponding to the
innermost regions of the halo. Of greater interest is the fact that
the estimated mass and the virial mass are uncorrelated. Given that
$r_s$ and $r_{vir}$ are uncorrelated, this is unsurprising, but it
means that simple mass estimates for these galaxy groups do not {\it
trace}, let alone {\it estimate}, the true mass of the dark matter
halo. We emphasise that the cause of this discrepancy is due to the
fact that, while the velocity dispersion of the group traces the
virial velocity of the group's dark matter halo, the scale radius of
the group is uncorrelated with the properties of the dark matter halo.

\section{Discussion}

In this paper we have explored the three dimensional spatial
properties of Hickson compact groups of galaxies identified in a mock
galaxy catalogue from \cite{blaizot2005}.  We now discuss our main
results in the context of previous observational studies.

\subsection{How compact are compact groups?}

The compactness of compact groups has been an issue of some
considerable debate. There have been many previous interpretations as
to what compact groups are, summarised in the Annual Review by
\cite{hickson1997}. Of particular relevance, \cite{mamon1986} argued that
roughly half of all HCGs were chance alignments of galaxies within
looser groups, a view echoed by \cite{walke1989} and \cite{mamon1995}.

We find that approximately 30\,\% of HAs in our mock catalogue are
physically dense systems of four or more galaxies with no
interlopers. Nearly one-quarter of HAs are not physically dense, even
in part. This latter fraction is significantly less than suggested by
\cite{mamon1986}, but it is still a considerable fraction. The
remaining HAs (48\,\%) are dense configurations of 2, 3 or 4 galaxies
with one or more interlopers. Thus the majority (over three-quarters)
of HAs are physically dense associations of galaxies, although most of
these contain interlopers and would otherwise fail to satisfy the
selection criteria for HAs were it not for the presence of these
interloping members. In terms of individual galaxies, 25\,497 of the
64\,525 galaxies identified by the Hickson criteria (40\,\%) are not
found in very close proximity to at least one other galaxy, and will
no doubt act to bias any study of galaxy properties in compact groups
(\citealt{brasseur2008}).

\cite{ponman1996} surveyed most of the HCGs with ROSAT. They
calculated that the X-ray emission they detected was greater than
would be expected from each galaxy in isolation, and implied that most
HCGs must therefore be physical associations of galaxies, with the
excess emission associated with hot inter-galactic gas in the
group. They implied that $\gtrsim 75\,\%$ of the HCGs were physical
associations of galaxies, by extrapolation of the number of
associations from which they measured detectable X-ray emission. We
find that $59\,\%$ of HAs are dense groupings of 3 or more galaxies,
and this figure rises to $77\%$ if pairs are included. If these
systems contain some intra-group hot gas, then our finding would be in
good agreement with the \cite{ponman1996} result.

\subsection{Are Hickson's criteria optimal?}

The identification of a small number of physically associated galaxies
based only on projected positions and apparent magnitudes is a
difficult task, and we have found that most HAs are compact groupings
of galaxies. However, only $29\,\%$ of HAs consist of 4 or more
galaxies in a compact configuration with no interlopers. The remaining
HAs can, in some respect, be viewed as contamination of this sample.

When we compare the observable properties of CAs with the
`contaminants' in Figure~5 and Table~1, we find that less populated,
brighter, more isolated HAs are more likely to be CAs than more
populated, fainter, and/or less isolated HAs. Table~\ref{t1} shows
that the success rate of identifying HA can be significantly improved
by changing the selection criteria. The best criterion to change in
this respect is the surface brightness limit; increasing the cut from
$\mu_e = 26$\,mags\,arcsec$^{-2}$ to $\mu_e = 23$\,mags\,arcsec$^{-2}$
increases the fraction of CAs identified from 29\,\% to 67\,\%. This
fraction can be increased to three-quarters by increasing the surface
brightness limit to $\mu_e = 22$\,mags\,arcsec$^{-2}$.  This is done
at a cost, of course, as fewer HAs identified with the original
criteria pass these more strict requirements. Based upon these
experiments, we conclude that, if you want to identify a sample of
compact groups with as few interlopers as possible, then you should
use selection criteria which only accept the brightest, most isolated,
compact groups.

\cite{lee2004} slightly modified the Hickson criteria
in their search for compact groups in the SDSS commissioning data,
such that groups had to have an effective surface brightness of $\mu_e
\le 24$\,mags\,arcsec$^{-2}$. This will have reduced the expected
contamination level to $\sim 44$\,\%, significantly less than would
have been obtained using the original criteria.

In a subsequent paper in this series we use the information presented
here to derive a large catalogue of CAs from the SDSS photometric
catalogue. Importantly, for the first time in a study of compact
groups identified in photometric catalogues, we will have a strong
understanding of our sample selection and can be statistically
confident in the identification of compact groups with minimal
contamination from interlopers.

\subsection{What are the dark matter properties of compact groups?}

We have selected compact groups based solely on the properties of the
luminous galaxies, independent of any dark matter properties. It is
therefore satisfying to see that, in making this selection, we
preferentially identify galaxies which belong to the same dark matter
halo. In over half the compact groups (55\%), all the member galaxies
are found within the same dark matter halo. A smaller number have two
or three haloes (32\% and 11\%, respectively), and only 2\,\% of
compact groups are made of galaxies which each have their own dark
matter halo.

We interpret the above result in terms of formation time. That is,
those compact groups where each galaxy has its own halo are
dynamically young, and the haloes have not yet had time to merge. In
contrast, those compact groups with only one halo are dynamically
older, and the haloes have fully merged.  However, the mock catalogs
used in this study do not contain information about the evolutionary
history of these galaxy groups and the dark matter haloes that they
occupy.  A proper treatment of the evolutionary properties of these
CAs would require analysis of the full de Lucia catalogs. As a
result, a more detailed treatment of this topic is beyond the scope of
this current contribution.

\cite{vennik1993}, \cite{rood1994} and \cite{ramella1994} showed that
the majority of the HCGs seemed to be associated with loose groups and
clusters, although these were not generally in the highest density
environments (eg. \citealt{sulentic1987,palumbo1995}). More recently,
an association between loose groups and a subset of the HCGs was also
found by \cite{tovmassian2006}. \cite{diaferio1994} and
\cite{governato1996} have both suggested that compact groups may arise
as dense configurations within looser groups. For those CAs where all
the galaxies reside within the same halo, we find that they generally
dominate the halo. However, in a significant proportion of cases,
there are several other galaxies in the halo which are not part of the
CA, implying that the CA is part of a looser group with typically 5 -
10 galaxies in total. The virial masses of these haloes are of order
$10^{13} - 10^{14}\,h^{-1}\,$M$_\odot$. Thus our results for both the
halo mass and the number of galaxies within the halo strongly suggest
that compact groups are associated with galaxy groups, some or all of
which have formed a dense galaxy grouping. This is in good agreement
with recent observational studies, and argues against compact groups
being associated with galaxy clusters.

Finally, we examine whether it is possible to relate the observed
dynamics of compact groups to the mass of the host halo. We find that
there is a strong statistical correlation between the host halo virial
velocity and the three dimensional velocity dispersion of the
group. However, the significant spread in this trend - likely a result
of the velocity dispersion being based on only four galaxies - means
that a significant uncertainty will be introduced if the velocity
dispersion is used as a proxy for the virial velocity for any
individual system. 

Our results indicate that compact groups are deeply embedded near the
centre of their dark matter halo. Unfortunately for observers, the
degree to which they are embedded is uncorrelated with the virial
radius of the halo. This result suggests a picture in which the
individual galaxies and their haloes merge together and the galaxies
sink towards the centre of their new host halo until they are within a
few hundred kiloparsecs of each other, at which point they can be
identified as a compact group.  Thus derived mass estimates of the
halo do not correlate with the true mass of the halo; by the time the
group is dense enough to be able to be identified as a compact group,
the linear size of the group - and hence the fraction of the halo mass
which the group samples - is independent of the properties of the
surrounding dark matter halo. In practice, this result will require
more precise modeling of observed compact group dynamics in order to
determine their dark matter halo properties.

\section*{Acknowledgments} 

The Millennium Simulation databases used in this paper and the web
application providing online access to them were constructed as part
of the activities of the German Astrophysical Virtual Observatory. AWM
acknowledges support from a Research Fellowship from the Royal
Commission for the Exhibition of 1851. He also thanks Sara Ellison and
Julio Navarro for additional financial assistance. We thank Jon
Willis, Crystal Brasseur and Jorge Pe{\~n}arrubia for useful and
stimulating discussions of these results.  SLE and DRP acknowledge the
receipt of NSERC Discovery Grants which funded some of this
research. We thank the anonymous referee for very thoughtful and useful
suggestions which led to significant improvements in this paper.

\bibliographystyle{apj}
\bibliography{/Users/Alan/Papers/references}

\end{document}